# A Physiological Control System for an Implantable Heart Pump that Accommodates for Interpatient and Intrapatient Variations

Masoud Fetanat*, Michael Stevens, Christopher Hayward, Nigel H. Lovell, *Fellow, IEEE*

**Abstract**— Left ventricular assist devices (LVADs) can provide mechanical support for a failing heart as bridge to transplant and destination therapy. Physiological control systems for LVADs should be designed to respond to changes in hemodynamic across a variety of clinical scenarios and patients by automatically adjusting the heart pump speed. In this study, a novel adaptive physiological control system for an implantable heart pump was developed to respond to interpatient and intrapatient variations to maintain the left-ventricle-end-diastolic-pressure (LVEDP) in the normal range of 3 to 15 mmHg to prevent ventricle suction and pulmonary congestion. A new algorithm was also developed to detect LVEDP from pressure sensor measurement in real-time mode. Model free adaptive control (MFAC) was employed to control the pump speed via simulation of 100 different patient conditions in each of six different patient scenarios, and compared to standard PID control. Controller performance was tracked using the sum of the absolute error (SAE) between the desired and measured LVEDP. The lower SAE on control tracking performance means the measured LVEDP follows the desired LVEDP faster and with less amplitude oscillations preventing ventricle suction and pulmonary congestion (mean and standard deviation of SAE(mmHg) for all 600 simulations were 18813±29345 and 24794±28380 corresponding to MFAC and PID controller respectively). In four out of six patient scenarios, MFAC control tracking performance was better than the PID controller. This study shows the control performance can be guaranteed across different patients and conditions when using MFAC over PID control, which is a step towards clinical acceptance of these systems.

*Masoud Fetanat (correspondence e-mail:m.fetanat@ieee.org), Michael Stevens and Nigel Lovell are with the Graduate School of Biomedical Engineering, University of New South Wales, Sydney, Australia. Christopher Hayward is with Cardiology Department, St Vincent's Hospital, Sydney, New South Wales, Australia; Victor Chang Cardiac Research Institute, Sydney, New South Wales, Australia; and School of Medicine, University of New South Wales, Sydney, New South Wales, Australia.

## I. INTRODUCTION

LEFT ventricular assist devices (LVADs), which are mechanical pumps implanted in patients with heart failure, have been used as a mechanical circulatory support to assist a failing left ventricle during bridge to transplant, bridge to recovery and destination therapy [1]. Operating LVADs at constant speed may lead to ventricular suction (ventricular collapse due to low pressure in the ventricle) or pulmonary congestion (a condition caused by excess fluid in the lungs due to high pressure in the ventricle). Ventricular suction may lead to hemolysis, heart tissue damage near the pump inlet, right ventricular dysfunction or release of ventricular thrombus and subsequent stroke, while pulmonary congestion may lead to flooding of the lungs (pulmonary edema) and shortness of breath [2]–[4].

To prevent these hazardous events, physiological control systems for LVADs can be used to automatically adjust pump speed in response to changes in hemodynamic or pump variables to satisfy control objective(s). These objectives may include a constant differential pressure [5], a constant pump inlet pressure [6], a constant pump flow [7] and a Starling-like controller which sets target flow rate based on the left ventricular end-diastolic pressure (LVEDP) [8]–[12]. Most investigators have previously employed proportional-integral-derivative (PID) controller, with fixed gains, to implement these physiological control systems [10]–[13].

These systems are commonly evaluated using computer simulations. However, in most of these simulations, [10]–[15], only a single patient and/or condition is simulated. This is not clinically representative of the wider interpatient and intrapatient variations in cardiovascular system (CVS) dynamics. The control system, usually PID control, is tuned to provide optimal performance only for this specific patient and condition. The consequence of this is that control performance cannot be guaranteed across different patients and conditions, which may lead to more hazardous events like suction and congestion.

An adaptive control system can adjust its parameters in the face of disturbances and time-varying plants. Adaptive control can be beneficial for controlling VADs because of the time-varying nature of the CVS. In a recent study, a neural predictive controller, which is one kind of adaptive controller, was employed to control dual rotary blood VADs by using a trained Artificial Neural Network (ANN) [15], and it was shown to have improvements over non-adaptive control. However, ANN controllers need training data which is difficult and time-consuming to collect. Furthermore, ANN controllers have a heavy computational burden which is not ideal from a practical viewpoint. Other non-linear adaptive controllers, such as fuzzy logic control (FLC), require the determination of rulesets which can be time consuming to produce and result in inefficient performance due to the non-linear behaviour [16], [17].

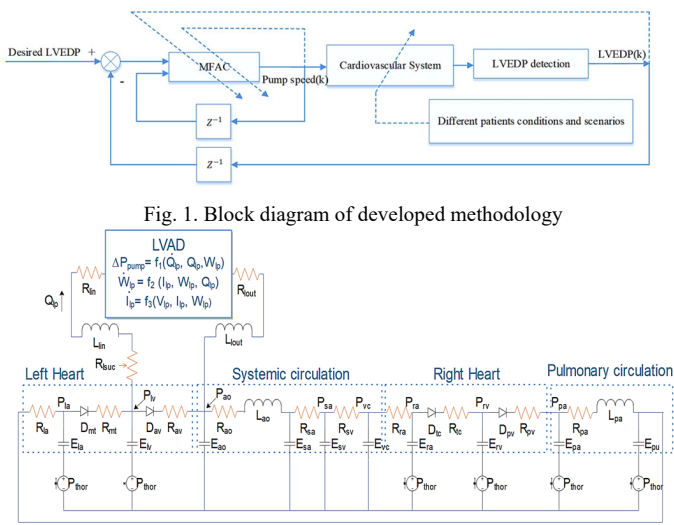

Fig. 1. Block diagram of developed methodology

Fig. 2. Electrical equivalent circuit analogue of cardiovascular with heart–pump interaction model. P, pressures; R, resistances; E, elastances (=1/compliances); L, inertances; D, diodes. The model consists of two main components: (1) the cardiovascular model includes left heart, systemic circulation, right heart, pulmonary circulation (la, left atrium; lv, left ventricle, ao, aorta; sa, systemic peripheral vessels, including the arteries and capillaries; sv, systemic veins, including small and large veins; vc, vena cava; ra, right atrium; rv, right ventricle; pa, pulmonary peripheral vessels, including pulmonary arteries and capillaries; pu, pulmonary veins and (2) heart-pump model includes the LVAD and RVAD, (Rin and Rout , inlet and outlet cannulae resistances; Lin and Lout, inlet and outlet cannulae inertances; Rlsuc, left suction resistance, Rrsuc, right suction resistance, Rband, banding resistance). The intrathoracic pressure, Pthor was assigned –4 mmHg during closed-chest simulated conditions [15], [18].

To address these issues, we propose the use of model free adaptive control (MFAC) to implement physiological control systems for rotary LVADs. MFAC does not require the training process like an ANN controller, nor does it require knowledge about the dynamic and structural information of the controlled system. It only uses the real-time measurement data of the controlled system, resulting in a generic controller for a class of industrial practical systems and leading to more efficient control. Furthermore, unlike the FLC, MFAC also does not require any rules, which makes it much easier to implement.
This study aims to develop an adaptive physiological controller that utilizes MFAC for an implantable LVAD that provides consistent control system performance across a range of interpatient and intrapatient variations. The objective of this controller is to minimize the error between the desired and measured LVEDP, and reduce the risk of hazardous events such as ventricle suction and pulmonary congestion.

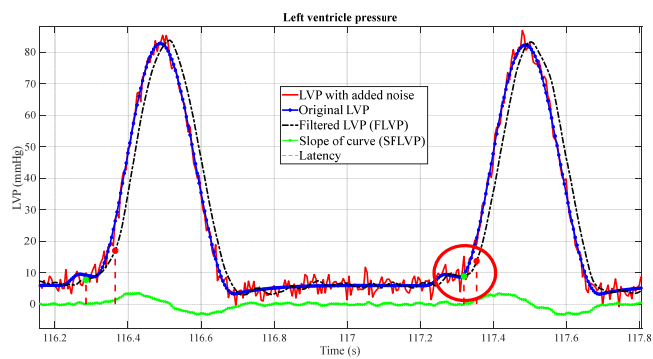

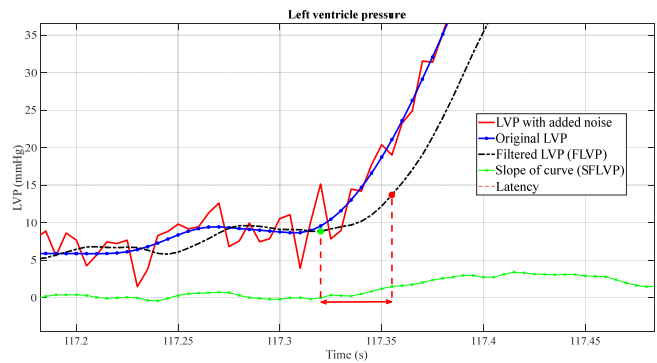

Fig. 3. Real-time LVEDP detection, (a) left ventricle pressure (LVP) in transition from rest to exercise for patient 1 with white noise variance of 2 mmHg, (b) zoomed LVP.

In the fo llowing sections, first the numerical model of the human CVS and heart pump adopted for the evaluation of the proposed physiological control method is introduced. Then, a novel method for detecting LVEDP robust to noise is presented. Afterwards, the MFAC is described. Then different patient conditions and scenarios used for evaluation of the MFAC are described. Subsequently, the sensitivity analysis (used to identify parameters from the CVS that most likely contribute to control performance variation) is then described. Then, the result of the sim ulations with interpatient and intrapatient variations is presented. Finally, a discussion according to the simulation results, limitations of the study and future works is presented.

## II. METHODOLOGY

Fig. 1 shows the block diagram of the control system and testing methodology used in this study. It consists of four major blocks: Numerical model of the cardiovascular system, LVEDP detection, MFAC system and evaluation using different patient conditions and scenarios. All are described in this section as follows.

### A. Numerical Model of Cardiovascular System

In this study, the simulations to investigate the implementation of the proposed LVEDP detection and MFAC

controller, implemented in Simulink (MathWorks, Natick, MA, USA), were carried out on the numerical model of the cardiovascular system developed by Lim et al. [18] (Fig. 2). The model was derived based on first principles and includes left and right hearts, systemic and pulmonary circulations, two heart pump models (LVAD and RVAD) and inlet and outlet cannulae. The CVS and heart pump model have been validated using in-vitro mock-loop tests and in-vivo animal experiments [18]–[20].

The VentrAssist$^{TM}$ centrifugal pump was used as the LVAD in this study (Ventracor Ltd., Sydney, Australia). The VentrAssist model used was based on the model provided in [19]–[21]. The speed range of the pump model was constrained between 1800 and 3000 rpm, the clinical operating range of the device.

### B. Real-time LVEDP Detection

As there is no method of LVEDP detection from LVP measurement in the literature, a real-time LVEDP detection method is described in this section. In this control system, LVEDP was used as the feedback variable because of its excellent performance in previous physiological control studies [13]. It is an important clinical variable indicative of ventricular performance and can identify clinical symptoms of heart failure in patients. It also has a much simpler feedback pathway than more complex Frank-Starling control systems. However, it is difficult to determine LVEDP accurately from pressure sensor measurement in real time due to sensor noise and a lack of ECG signal to trigger LVEDP identification. To address this, we developed a novel LVEDP detection method that has 9 steps:

1. Smoothing the left ventricle pressure using a Butterworth low-pass filter with pass frequency of 5 Hz and stop frequency of 20 Hz named as filtered LVP (FLVP).
2. Computing the slope of the FLVP (SFLVP).
3. Computing heart beat based on the finding the time difference between two consecutive peaks from SFLVP.
4. Computing the mean of SFLVP within the last window of 10 samples (MSFLVP).
5. Finding the points which meet FLVP $\geq \alpha$ MSFLVP
6. Computing an adaptive threshold (TH) equal to the mean of 15 highest values in the SFLVP from previous cardiac cycle to current time, multiplied by $\beta$.
7. Finding the first point meeting the step 5 and 6 at the same time (LVEDP detection time).
8. Finding the closest point between the local minimum of the SFLVP and the LVEDP detection time found in step 7 (LVEDP actual time).
9. Computing the LVEDP from FLVP at LVEDP actual time found in step 8.

where $\alpha$ and $\beta$ were scaling factors, empirically adjusted to give a latency of 30 ms and sum of the absolute error of 1.2 mmHg when a noise signal with variance of 4 mmHg was added to the left ventricular pressure.

Fig. 3 (a) shows the LVP in transition from rest to exercise for a simulated patient with white noise variance of 2 mmHg.

Fig. 3 (b) depicts the original LVP, LVP with added noise, filtered LVP (FLVP), slope of the FLVP (SFLVP), identified

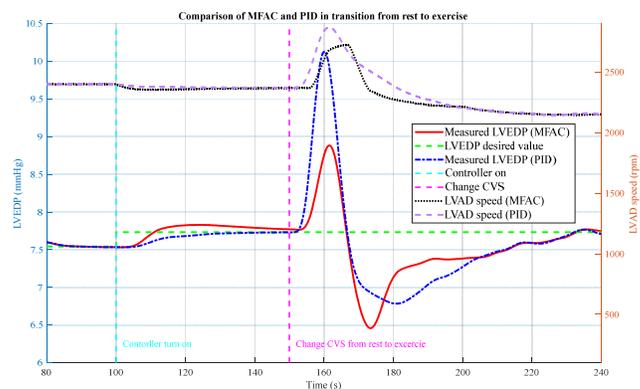

Fig. 4. A sample simulation for comparison of MFAC and PID in transition from rest to exercise for patient 45.

TABLE I
NORMAL VALUES AND DESCRIPTION OF CARDIOVASCULAR MODEL PARAMETERS FOR INTERPATIENT AND INTRAPATIENT SIMULATIONS

| No | Parameter (unit) | Description | Normal Value |
|---|---|---|---|
| 1 | Eeslvf (mmHg.mL$^{-1}$) | LV end systolic elastance | 3.54 |
| 2 | Eesrvf (mmHg.mL$^{-1}$) | RV end systolic elastance | 1.75 |
| 3 | Eao (mmHg.mL$^{-1}$) | Aortic elastance | 1.04 |
| 4 | Eesla (mmHg.mL$^{-1}$) | LA end systolic elastance | 0.2 |
| 5 | Eesra (mmHg.mL$^{-1}$) | RA end systolic elastance | 0.2 |
| 6 | Epa (mmHg.mL$^{-1}$) | Pulmonary arterial elastance | 0.15 |
| 7 | Epu (mmHg.mL$^{-1}$) | Pulmonary vein elastance | 0.04 |
| 8 | Esa (mmHg.mL$^{-1}$) | Systemic arterial elastance | 0.37 |
| 9 | Esv (mmHg.mL$^{-1}$) | Systemic vein elastance | 0.013 |
| 10 | Evc (mmHg.mL$^{-1}$) | Vena cava elastance | 0.03 |
| 11 | Rao (mmHg.s.mL$^{-1}$) | Aortic resistance | 0.2 |
| 12 | Rra (mmHg.s.mL$^{-1}$) | Right atrium resistance | 0.012 |
| 13 | Rpv (mmHg.s.mL$^{-1}$) | Pulmonary valve resistance | 0.02 |
| 14 | Rsv (mmHg.s.mL$^{-1}$) | Systemic venous resistance | 0.12 |
| 15 | Tc (s) | Heart rate coefficient | 1 |
| 16 | Tsys0 (s) | Maximum systolic heart period | 0.5 |
| 17 | V0la (mL) | LA end diastolic volume at zero pressure | 20 |
| 18 | V0lvf (mL) | LV end diastolic volume at zero pressure | 40 |
| 19 | V0ra (mL) | RA end diastolic volume at zero pressure | 20 |
| 20 | V0rvf (mL) | RV end diastolic volume at zero pressure | 50 |
| 21 | Vdla (mL) | LA end systolic volume at zero pressure | 10 |
| 22 | Vdlvf (mL) | LV end systolic volume at zero pressure | 16.77 |
| 23 | Vdra (mL) | RA end systolic volume at zero pressure | 10 |
| 24 | Vdrvf (mL) | RV end systolic volume at zero pressure | 40 |
| 25 | Rmt (mmHg.s.mL$^{-1}$) | Mitral valve resistance | 0.01 |
| 26 | Rav (mmHg.s.mL$^{-1}$) | Aortic valve resistance | 0.02 |
| 27 | Vuao (mL) | Aortic unstressed volume | 230.88 |
| 28 | Vupa (mL) | Pulmonary arterial unstressed volume | 91.67 |
| 29 | Vupu (mL) | Pulmonary vein unstressed volume | 132.39 |
| 30 | Vusa (mL) | Systemic arterial unstressed volume | 231.04 |
| 31 | Vusv (mL) | Systemic vein unstressed volume | 1976.1 |
| 32 | Vuvc (mL) | Vena cava unstressed volume | 136.17 |
| 33 | P0la (mmHg) | LA end diastolic stiffness scaling term | 0.5 |
| 34 | P0lvf (mmHg) | LV end diastolic stiffness scaling term | 0.98 |
| 35 | P0ra (mmHg) | RA end diastolic stiffness scaling term | 0.5 |
| 36 | P0rvf (mmHg) | RV end diastolic stiffness scaling term | 0.91 |
| 37 | Vtotal (mL) | Total blood volume | 5200 |
| 38 | $\lambda$la (mL$^{-1}$) | LA end diastolic stiffness coefficient | 0.025 |
| 39 | $\lambda$lvf (mL$^{-1}$) | LV end diastolic stiffness coefficient | 0.028 |
| 40 | $\lambda$ra (mL$^{-1}$) | RA end diastolic stiffness coefficient | 0.025 |
| 41 | $\lambda$rvf (mL$^{-1}$) | RV end diastolic stiffness coefficient | 0.028 |
| 42 | Lao (mmHg.s$^2$.ml$^{-2}$) | Aortic inertance | 0.0001 |
| 43 | Lpa (mmHg.s$^2$.ml$^{-2}$) | Pulmonary arterial inertance | 7.70e-05 |

LVEDP and latency between LVEDP detection time and LVEDP actual time. As it can be seen in the Fig. 3 (b), the green point shows the exact location of LVEDP before ventricle contraction. The sampling rate of measured variables is 200 Hz.

The LVEDP detection method was evaluated in six different patient scenarios; rising and falling pulmonary vascular resistance (PVR), rising and falling systemic vascular resistance (SVR), rest to exercise and postural change with 100 different patient conditions in each. To vary the SVR and PVR in the model, systemic arterial resistance ($R_{sa}$) and pulmonary arterial resistance ($R_{pa}$) were changed, as arterial resistance is the main contributor to total vascular resistance in the numerical model. In order to evaluate the impact of the pressure sensor noise on the developed LVEDP detection, a white noise signal with different levels of variance (0 to 4) was added to the left ventricular pressure. Mean of absolute error between LVEDP detection and actual time is defined by (1):

$$MAE = \frac{1}{n}\sum_{i=1}^{n} |LVEDP\ detection\ time_i - LVEDP\ actual\ time_i| \quad (1)$$

where *MAE* is the mean of the absolute error (millisecond), *LVEDP detection time$_i$* and *LVEDP actual time$_i$* are detection and real locations (time) of the LVEDP derived from the left ventricle pressure and n is the number of total samples in each simulation which equals to the number of cardiac cycles in each scenario calculated by LVEDP detection method.

### C. Model Free Adaptive Control (MFAC)

The MFAC is an adaptive control system which does not need the model of the plant to design the controller. It is designed by using the input/output measurement data of the controlled plant, without knowledge of the system structure or dynamics information of the controlled plant explicitly or implicitly. Specifically, MFAC identifies the linear or nonlinear model of the plant via a series of equivalent dynamic linearization data models within dynamic operation points of the closed-loop system by using a dynamic linearization technique (DLT) and pseudo-partial derivative (PPD) [16], [17]. A brief description of the MFAC follows.

A single input, single output (SISO) nonlinear discrete-time system can be defined as follows:

$$y(k+1) = f(y(k), \dots, y(k-n_y), u(k), \dots, u(k-n_u)) \quad (2)$$

where $y(k) \in R$ and $u(k) \in R$ are the system output and control input at time $k$, $n_y$ and $n_u$ are unknown orders of output and input, and $f(\dots) \in R$ is the unknown nonlinear function [16].

The compact form dynamic linearization of nonlinear system (2) can be defined by (3):

$$y(k+1) = y(k) + \varphi(k)\,\Delta u(k) \quad (3)$$

where $\varphi(k)$ is called pseudo-partial derivative (PPD).

In order to find the control input u(k), it is considered to minimize the following equation to have less error (difference between the desired and system output) and changes of two consecutive control inputs:

$$J(u(k)) = \|y^*(k+1) - y(k+1)\|^2 + \lambda \|u(k) - u(k-1)\|^2 \quad (4)$$

where $y^*(k+1)$ and $\lambda$ are desired output and a weighting constant respectively [16].

By substituting (3) into (4) and differentiating (4) with respect to $u(k)$ equals to zero, the control input is as follows:

$$u(k) = u(k-1) + \frac{\rho\,\varphi(k)\,(y^*(k+1) - y(k))}{\lambda + \|\varphi(k)\|^2} \quad (5)$$

where $\rho$ is a step constant which is added to make (5) more general and flexible.

As the unknown PPD parameter $\varphi(k)$ is time-varying, the least squares algorithm cannot track it well. Therefore, a time-varying estimation algorithm is used to estimate PPD $\varphi(k)$. By using the modified projection algorithm, the estimated unknown PPD $\varphi(k)$ can be found as follows [22]:

The objective function for $\varphi(k)$ estimation is defined as in (6)

$$J(\hat{\varphi}(k)) = \|\Delta y(k) - \hat{\varphi}(k)\,\Delta u(k-1)\|^2 + \mu \|\hat{\varphi}(k) - \hat{\varphi}(k-1)\|^2 \quad (6)$$

where $\mu > 0$ is a weighing factor.

By minimizing the objective function (6) with respect to $\hat{\varphi}(k)$, estimated unknown PPD ($\hat{\varphi}(k)$) is found as follows:

$$\hat{\varphi}(k) = \hat{\varphi}(k-1) + \frac{\eta\,\Delta u(k-1)\,(\Delta y(k) - \hat{\varphi}(k-1)\,\Delta u(k-1))}{\mu + \|\Delta u(k-1)\|^2} \quad (7)$$

$$\hat{\varphi}(k) = \hat{\varphi}(1), if\ |\hat{\varphi}(k)| \le \varepsilon\ or\ \Delta u(k-1) \le \varepsilon \\ or\ sign(\hat{\varphi}(k)) \ne sign(\hat{\varphi}(1)) \quad (8)$$

where $\eta \in (0,1]$ is a step constant which is added to make (7) more general and flexible, $\hat{\varphi}(1)$ is the initial value of $\hat{\varphi}(k)$ and $\varepsilon$ is a very small positive constant. The reset algorithm (8) is employed to make a stronger ability in estimating the time-varying parameter by the parameter estimation algorithm [16]. MFAC is dependent on following four assumptions:

Assumption 1: The partial derivative of $f(\dots)$ in (2) with respect to the $(n_y + 2)th$ variable is continuous.

Assumption 2: (2) satisfies the generalized Lipschitz condition.

Assumption 3: For a given bounded desired output signal $y^*(k+1)$, there exists a bounded control input $u^*(k)$ such that the system output reaches to the $y^*(k+1)$.

Assumption 4: The sign of PPD $\hat{\varphi}(k)$ is assumed to unchanged.

Nonlinear system (2) satisfying assumption 1, 2, 3 and 4 is controlled by (5), (7) and (8) for a regulation problem; i.e., $y^*(k+1) = y^* = const$, then there exists a constant $\lambda_{min}$ such that [17]:

1. $u(k)$ and $y(k)$ are bounded and the closed loop system is bounded input, bounded output (BIBO) stable.
2. The tracking error converges monotonically; i.e., $\lim_{k \to \infty} |y^* - y(k+1)| = 0$

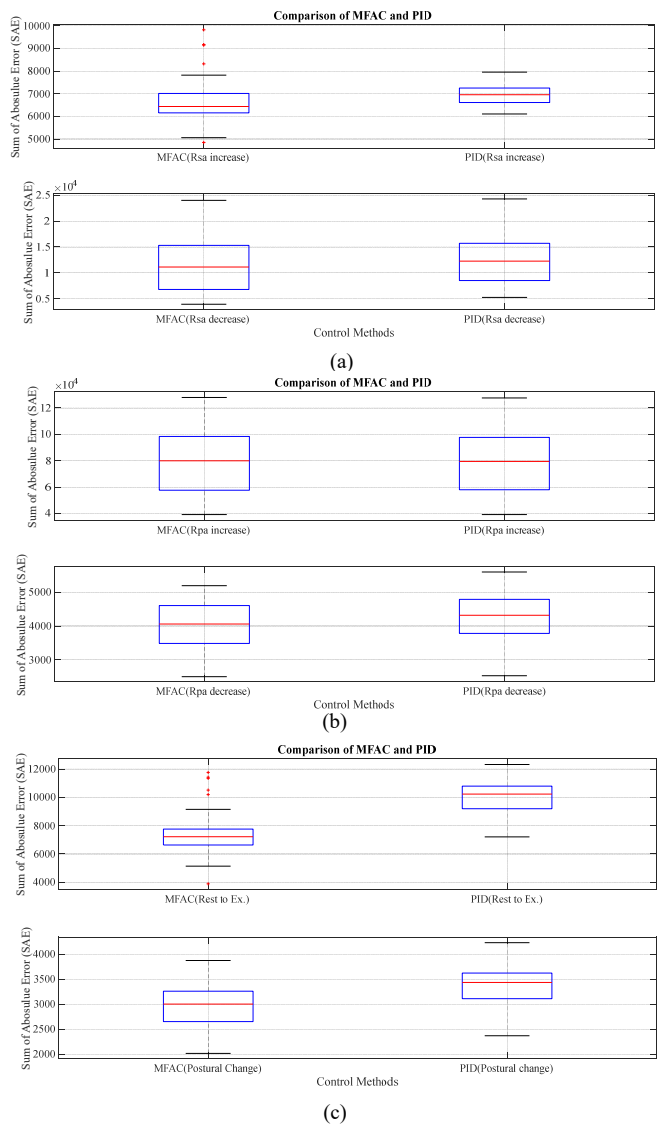

Fig. 5. Comparison of SAE of MFAC and PID controller in different scenarios via box plot, the central mark (red line) shows the median, and the bottom and top edges (blue lines) of each box show the 25th and 75th percentiles. The whiskers extend to the most extreme data points not considered outliers, and the outliers are plotted individually using the '+' symbol. (a) rising $R_{sa}$ (1300 to 600 dyne.s.cm-5) and falling $R_{sa}$ (1300 to 2600 dyne.s.cm-5) (b) rising $R_{pa}$ (100 to 40 dyne.s.cm-5) and falling $R_{pa}$ (100 to 500 dyne.s.cm-5) (c) transition from rest to exercise and passive postural change.

### D. Different patient conditions and scenarios

The MFAC controller was subject to a range of inter- and intrapatient variations for evaluation. In order to simulate these variations, some of the cardiovascular model parameters from TABLE I were changed from -20% to +20% of their nominal values to create 100 different patient conditions in each of six patient scenarios. To identify which parameters to vary, sensitivity analysis was used to find which model parameters had the greatest effect on the control tracking performance, defined in (9), in each patient scenario. The tracking performance derived under PID control was calculated in each simulation.

TABLE II
SIX DIFFERENT PATIENT SCENARIOS AND THEIR SIMULATION CONDITIONS

| Patient scenarios | Simulation conditions | Reference |
|---|---|---|
| Increasing $R_{pa}$ changes | 100 to 500 dyne.s.cm$^{-5}$ | [13] |
| Decreasing $R_{pa}$ changes | 100 to 40 dyne.s.cm$^{-5}$ | [13] |
| Increasing $R_{sa}$ changes | 1300 to 2600 dyne.s.cm$^{-5}$ | [13] |
| Decreasing $R_{sa}$ changes | 1300 to 600 dyne.s.cm$^{-5}$ | [13] |
| Transition from rest to exercise | 1. Increasing heart rate from 60 to 80 bpm<br>2. Decreasing $R_{pa}$ from 100 to 40 dyne.s.cm$^{-5}$<br>3. Decreasing $R_{sa}$ from 1300 to 670 dyne.s.cm$^{-5}$<br>4. Adding 500 mL fluid from a reservoir into the heart via the right atrium | [13] |
| Passive postural change | Removing 300 mL fluid from the heart into a reservoir | [13] |

TABLE III
SIMULATION OF INTERPATIENT AND INTRAPATIENT VARIATIONS VIA CHANGING THE MOST EFFECTIVE PARAMETERS IN EACH SCENARIO

| Patient scenarios | Most effective parameters |
|---|---|
| $R_{pa}$ increase | Eesrvf, Esa, Esv, Evc, Rsv, Rmt, Vusv, Vupu, Vusa, Vusv, Vuvc, Vtotal, λlvf and λrvf |
| $R_{pa}$ decrease | Eesrvf, Eesra, Esv, Rsv, Vusv, Vtotal, λra and λrvf |
| $R_{sa}$ increase | Esa, Vusv, Vtotal and λrvf |
| $R_{sa}$ decrease | Eeslvf, Eesrvf, Eesra, Esa, Esv, Evc, Rsv, V0lvf, Vdlvf, Rmt, Vusv, P0lvf, P0rvf, Vtotal, λlvf, λra and λrvf |
| Transition from rest to exercise | Eeslvf, Eesrvf, Eesla, Eesra, Esv, Evc, Rsv, V0lvf, Rmt, Vusv, Vtotal, λla, λra and λrvf |
| Passive postural change | Eeslvf, Eesrvf, Esv, Evc, Rsv, V0lvf, Vdlvf, Rmt, Vusv, P0lvf, Vtotal, λla, λlvf and λrvf |

$$SAE = \sum_{i=1}^{n} |LVEDP_d - LVEDP_m| \quad (9)$$

where $LVEDP_d$ is desired output (target), $LVEDP_m$ is model output (measured output), SAE is the sum of absolute error and $n$ is number of samples.

Afterward, dimensionless parameter sensitivity coefficient $S_j$ was evaluated in each of the six patient scenarios using (10).

$$S_j = \frac{\theta_j}{F_0} \frac{\Delta F}{\Delta \theta_j} \quad (10)$$

where $F_0$ is the initial nominal value of the objective function in (9), $\theta_j$ is the value of parameter from its baseline, $\Delta F$ is the difference between objective function derived by perturbed parameter and $F_0$, and $\Delta \theta_j$ is the difference between value perturbed parameter and $\theta_j$. The higher the sensitivity coefficients $S_j$, the greater the effect that parameter has on control tracking performance.

Sensitivity analysis was performed for the six common patient scenarios used to evaluate the MFAC. These scenarios were simulated by changing vascular resistance, heart rate and reservoir volume. These scenarios were rapid $R_{pa}$ changes (increasing and decreasing), rapid $R_{sa}$ changes (increasing and decreasing), transition from rest to exercise and a passive postural change (similar to a head-up tilt) experiment [13]. The simulation conditions for each scenario are shown in TABLE II. In transition from rest to exercise and passive postural change scenarios, each parameter is varied using a first order system with a time constant of 10 (s) which simulates the response of the native circulatory system. The other scenarios are simulated to change the $R_{pa}$ and $R_{sa}$ in rapid mode (step change). If a controller can respond properly to these extreme scenarios, it is assumed the controller can also respond to mild changes appropriately [13].

In each scenario, parameters were deemed significant if the sum of the sensitivity coefficient for both +20% and -20% parameter variations was at least 0.45.

### E. MFAC Evaluation

In order to evaluate the PID and MFAC reponses across interpatient and intrapatient variations, a suite of simulated LVAD patients were created for each of the six scenarios (totaling 600 "patients"). To create each patient, the most significant of the 43 cardiovascular model parameters (as determined by the results of the sensitivity analysis) were randomly changed between -20% to 20% from their nominal values.

In each patient for each scenario, an MFAC control system was developed to maintain LVEDP at a constant value, according to the following protocol. In the first 100 seconds, the heart pump works in a constant speed mode (2400 rpm) which allows the cardiovascular system to reach steady state. From time 100 sec, the MFAC was activated. At this point, if the measured LVEDP is between 3 mmHg to 15 mmHg, the initial set point is chosen as 0.2 mmHg greater the current value of LVEDP. This enables a simple step response to be evaluated. 150 seconds after commencement, one of the six patient scenarios were simulated. This process was repeated for a system controlled with PID control. PID gains with an anti-windup mechanism were tuned using a quasi-Newtonian optimization algorithm to minimize (9) as the same approach provided in [11]. The PID parameters were $k_p = 133.09$, $k_i = 17.17$, $k_d = 10.21$. The MFAC parameters were chosen empirically as $\mu = 0.1$, $\lambda = 0.1$, $\rho = 1$, $\eta = 1$, $\varphi(1) = 0.001$ and $\varepsilon = 10^{-4}$.

To assess the difference between PID and MFAC performance, the sum of absolute error between desired LVEDP and measured LVEDP defined in (9) was employed to compare the two controllers. Wilcoxon's test was performed to compare the SAE of PID and MFAC across the 100 simulated patients for each scenario, with a p-value less than 0.05 considered significant. Additionally, the number of scenarios in which pulmonary congestion occurred was noted. Based on the normal range of measured LVEDP (3 to 15 mmHg) [23], pulmonary congestions was defined the measured LVEDP above 15 mmHg. All values are expressed as mean ± standard deviation unless otherwise stated.

## III. RESULTS

### A. LVEDP Detection

TABLE IV shows the result of the developed real-time LVEDP detection in six different patient scenarios (rising and falling $R_{pa}$, rising and falling $R_{sa}$, rest to exercise and postural change) for 100 different patient conditions in each case. The results show that in all scenarios (600 simulations) with different levels of white noise, the developed LVEDP detection method has a very small latency (31.62±12.67 ms) for all simulations. Accuracy of LVEDP detection method was assigned by calculating of mean and standard deviation of the absolute error between the actual and detected LVEDP with different level of noises in each scenario shown in TABLE IV. The accuracy of the developed method for all the simulations is 0.71±0.61 mmHg.

TABLE IV
EVALUATION OF LVEDP DETECTION

| Scenarios | Total number of simulations in each scenario (number of cardiac cycles in each simulation) | White noise variance (mean of SNR) | Accuracy of developed method (mean ± std) | Latency of developed method ms (mean ± std) |
|---|---|---|---|---|
| Rising $R_{pa}$ | 100 different patient conditions (190) | 0 (NaN) | 0.32 ± 0.10 | 31.11 ± 4.98 |
| | | 1 (30.11) | 0.38 ± 0.34 | 32.65 ± 9.87 |
| | | 2 (24.11) | 0.64 ± 0.55 | 33.20 ± 14.61 |
| | | 3 (20.60) | 0.93 ± 0.75 | 32.03 ± 15.76 |
| | | 4 (18.14) | 1.27 ± 0.98 | 30.69 ± 16.45 |
| Falling $R_{pa}$ | 100 different patient conditions (190) | 0 (NaN) | 0.36 ± 0.08 | 30.02 ± 0.36 |
| | | 1 (31.53) | 0.41 ± 0.32 | 32.31 ± 7.66 |
| | | 2 (25.52) | 0.63 ± 0.54 | 33.86 ± 13.19 |
| | | 3 (22.01) | 0.91 ± 0.73 | 33.65 ± 15.20 |
| | | 4 (19.53) | 1.22 ± 0.96 | 32.60 ± 16.45 |
| Rising $R_{sa}$ | 100 different patient conditions (190) | 0 (NaN) | 0.39 ± 0.31 | 29.24 ± 6.78 |
| | | 1 (30.31) | 0.41 ± 0.44 | 30.99 ± 9.80 |
| | | 2 (24.32) | 0.62 ± 0.59 | 31.98 ± 14.20 |
| | | 3 (20.84) | 0.89 ± 0.77 | 31.74 ± 16.16 |
| | | 4 (18.40) | 1.18 ± 0.97 | 30.76 ± 17.00 |
| Falling $R_{sa}$ | 100 different patient conditions (190) | 0 (NaN) | 0.39 ± 0.07 | 29.02 ± 2.57 |
| | | 1 (30.67) | 0.40 ± 0.31 | 30.74 ± 8.18 |
| | | 2 (24.66) | 0.62 ± 0.50 | 32.16 ± 13.11 |
| | | 3 (21.16) | 0.90 ± 0.72 | 31.66 ± 14.79 |
| | | 4 (18.68) | 1.23 ± 0.97 | 30.75 ± 15.83 |
| Rest to Exercise | 100 different patient conditions (217) | 0 (NaN) | 0.41 ± 0.08 | 29.04 ± 2.15 |
| | | 1 (31.23) | 0.44 ± 0.40 | 31.27 ± 10.23 |
| | | 2 (25.22) | 0.67 ± 0.63 | 32.83 ± 15.06 |
| | | 3 (21.71) | 0.94 ± 0.81 | 32.56 ± 16.37 |
| | | 4 (19.24) | 1.26 ± 1.02 | 31.41 ± 16.75 |
| Postural Change | 100 different patient conditions (190) | 0 (NaN) | 0.38 ± 0.08 | 30.36 ± 2.36 |
| | | 1 (31.34) | 0.40 ± 0.30 | 32.18 ± 7.66 |
| | | 2 (25.33) | 0.63 ± 0.54 | 33.43 ± 12.89 |
| | | 3 (21.83) | 0.90 ± 0.72 | 33.05 ± 14.85 |
| | | 4 (19.35) | 1.22 ± 0.97 | 32.07 ± 15.98 |

### B. Sensitivity Analysis

The result of the sensitivity analysis performed by variation of each 43 parameters in six patient scenarios (TABLE III) shows that the cardiovascular model parameters total blood volume ($V_{total}$), systemic vein unstressed volume ($V_{usv}$), RV end diastolic stiffness coefficient ($\lambda_{rvf}$), systemic vein elastance ($E_{sv}$), RV end systolic elastance ($E_{esrvf}$), systemic vein resistance ($R_{sv}$) have the most impact on the control tracking performance in all the patient scenarios (whole results omitted due to space limitations on the manuscript but was added as supplementary materials in TABLE VI). These parameters were therefore varied to simulate the 100 patients for each scenario.

### C. MFAC Evaluation

As can be seen from Fig. 4, MFAC responds faster to patient changes, resulting in a smaller SAE than PID (solid red line compared to dash-dot blue line). Futhermore, in three simulations out of 100 simulations in transition from rest to exercise, PID conroller led to transient pulmonary congestions; however, MFAC could prevent pulmonary congestion due to the lower peak on the measured LVEDP in those three simulations.

Fig. 5 shows the comparison of SAE of MFAC and PID control for all patients in all scenarios. From the figure, in four patient scenarios (decreasing $R_{pa}$, increasing $R_{sa}$, rest to exercise and postural change), SAE (mmHg) derived from MFAC are much less than SAE from PID control (3989.48±700.84 and 5869.55±1157.66 for $R_{pa}$ decrease, and 6608.90±830.94 and 10841.26±1745.66 for $R_{sa}$ increase, 7353.51±1165.62 and 16948.92±4104.39 for rest to exercise, and 2945.74±403.39 and 3330.475±428.37 for postural change corresponding to MFAC and PID controller respectively). In two other scenarios (increasing $R_{pa}$ and decreasing $R_{sa}$) SAE (mmHg) derived from MFAC is almost similar or better than the PID controller (80270.31±23575.38 and 80087.31±23356.09 for $R_{pa}$ increase, and 11715.23±5161.54 and 31687.86±9183.17 for decrease $R_{sa}$ corresponding to MFAC and PID controller respectively). Mean and standard deviation of the SAE for all the 600 simulations are 18813.86±29345.78 and 24794.22 ± 28380.48 corresponding to MFAC and PID controller respectively.

TABLE V demonstrates the statistical result of the Wilcoxon's test in comparison of MFAC and PID controller for different patient conditions. In four out of six patient scenarios (decreasing $R_{pa}$, increasing $R_{sa}$, rest to exercise and postural change) the *p*-values are less than 0.05, indicating that MFAC has better performance than PID control.

## IV. DISCUSSION

One of the key issues of current physiological control systems [10]–[13], [15] is that they have been simulated in very specific conditions for single patient scenarios. This is not realistic, as each patient's cardiovascular system dynamics vary throughout the day and are unlikely to be identical to another patient. We argue that the control systems evaluated against a single simulated patient cannot be guaranteed to be robust against a wide patient cohort. The use of single-patient evaluation may also be another reason for the popularity of simple linear PID control in LVAD control systems [8]–[11]. It has yet to be established how simple PID control fares in the context of a wider patient suite.

In this study, we compared linear PID control and MFAC across a range 600 randomized patient scenarios. The results indicate that MFAC offers either similar or consistently better performance across all scenarios when compared to PID control. This indicates that the previous evaluation techniques were remiss in not considering the effect of plant variation on control system performance.

The results from Fig. 5 and TABLE V show that MFAC can improve the tracking performance compared to PID control since MFAC adjusts its parameters adaptively only based on the input/output data of the cardiovascular system. In fact, MFAC parameters were automatically changed by interpatient and intrapatient variations and any unpredicted changes on hemodynamics can be responded by MFAC due to the adaptive structure of the controller. There is also a mathematical proof that confirms the error between desired and measured output converges to zero by using the MFAC method. The lower SAE on control tracking performance means the measured LVEDP follows the desired LVEDP in the range of 3 to 15 mmHg faster and with less amplitude oscillations leading to fewer ventricular suctions and pulmonary congestion events. From the results it's clear that in the four patient scenarios (increasing $R_{sa}$, rest to exercise, decreasing $R_{pa}$ and postural change) MFAC control tracking performance is much better compared to PID controller, in the other two scenarios (increasing $R_{pa}$ and decreasing $R_{sa}$) SAE derived from MFAC is not significantly different than compared to PID. Optimization of finding the MFAC parameters can lead to further improvement in control tracking performance compared to the optimized PID controller.

TABLE V
RESULTS OF THE WILCOXON'S TEST IN COMPARISON OF MFAC AND PID FOR DIFFERENT PATIENT SCENARIOS

| Scenarios | *p*-value |
|---|---|
| Increasing $R_{pa}$ | 0.9542 |
| Decreasing $R_{pa}$ | **0.0198** |
| Increasing $R_{sa}$ | **1.7067e-06** |
| Decreasing $R_{sa}$ | 0.0998 |
| Rest to exercise | **9.9469e-28** |
| Postural change | **1.5741e-10** |

Another advantage of MFAC over PID is that tuning and optimization of PID controllers must be performed offline by a control engineer. If the performance is unsatisfactory, a control engineer familiar with the cardiovascular system may be required to make adjustments, as clinicians don't have the specialist engineering knowledge to make these adjustments. On the other hand, MFAC just uses the real-time measurement I/O data of the controlled system to self-adjust its parameters and so no additional input is required. Clinicians can therefore focus on treating their patient and not worry about underperforming or unstable control systems.

Although an adaptive physiological controller of rotary blood pumps using intrinsic pump parameters on two different pumps was proposed by Wu [24], it was based on a linear state-space model of cardiovascular system. This approach is difficult to use for simulation of different patient conditions, as the model linearization may not be constant across the suite of patients. This may lead to inaccuracy of simulation and therefore inconsistent control performance. Furthermore, the pump flow derived from feedback signals of intrinsic pump parameters was assumed to be the total flow to the human circulation system, which increases the estimation error of aortic pressure [24]. In addition, there was no methodology to prevent the possible suction and congestion events [24]. However, in this study, a stable model-free adaptive controller was employed to provide guaranteed controller performance across a range of 600 different patient conditions (six different scenarios), preventing suction and congestion events. In addition, the MFAC controller doesn't have the common issues of other model-free adaptive controllers like ANN and Fuzzy controllers (i.e. difficulty to collect training data and finding the rules). Model predictive controllers and optimal controllers require a simplified and generalized state-space model that cannot be modified for different patients' conditions.

As reported in the literature [10]–[12], Starling-like controllers have been proposed to restore Frank-Starling response of the native heart. One of the issues with the Frank-Starling mechanism is that the desired flow for Frank-Starling curve must be set manually by clinicians for each patient. As there is no automatic procedure on finding the Frank-Starling curve without clinical check-up, employing Starling-like controller for different patient conditions is impossible. In order to minimize clinical check-up and find an automatic procedure to set desired flow or LVEDP, constant LVEDP scheme was used in this study to maintain LVEDP in the range of 3 mmHg to 15 mmHg for each patient (to prevent ventricle suction and pulmonary congestion). This can be an appropriate way to respond to the interpatient and intrapatient variations.

The evaluation of the developed LVEDP detection in TABLE IV shows that the proposed method can identify the end-diastolic pressure across a range of different levels of noise with very low delay in a real-time mode. To the author's knowledge, there are no other LVEDP detection methods like this. Other investigators have previously addressed this by utilizing the minimum left ventricular pressure, which is not a realistic measure of preload in the ventricle [11]. The LVEDP detection method proposed in this paper will enable robust detection of LVEDP for physiological control, improving the performance of physiological control systems.

This study has some limitations. The first limitation is that our developed methodology hasn't employed the Frank-Starling mechanism to balance systemic and pulmonary flow. Frank-Starling-like controllers setting flow rate as a function of preload have been developed in [9], [19], and shown to be one of the best performing physiological control systems [13]. However, finding the Frank-Starling function for each different patient with various conditions is difficult. Therefore, to simplify the controller, LVEDP was maintained as a constant value in the range between 3 mmHg to 15 mmHg.

Second, the baroreflex was not simulated in this study which may influence performance and hemodynamics [25]. However, the MFAC controller can compensate its impact properly since any changes due to the baroreflex can be interpreted as a nonlinearity like to a different patient condition or scenario.

Third, this study assumes the left ventricular pressure can be measured. However, currently, there are no commercial long-term implantable pressure sensors. Alternatively, the pressure sensor can be replaced by pressure estimation algorithms which have been used in several original implementations of evaluated control systems like that for constant pump flow [7]. Further investigation is required to determine if the developed physiological control system can use pressure estimation algorithms instead of sensors [26]–[28].

Fourthly, this study was completed in-silico. Most LVAD physiological control systems are evaluated on the bench top using mock circulation loops, to verify numerical model findings. However, the numerical model used in this study was extensively validated using animal data. Additionally, the numerical model is easily scalable: it enabled us to easily simulate 600 patient conditions across 6 different scenarios. This is a laborious process to conduct using bench top testing.

Future work will involve validating this system on the bench top and in-vivo.

Fifthly, this study used a model of the VentrAssist LVAD. The VentrAssist is no longer used clinically, however as it is a centrifugal pump like the commonly used HeartWare HVAD (Medtronic Inc., FL, USA), it is hypothesized that there would be little difference in performance. Future work will involve repeating this experiment using models of clinically available pumps.

Finally, pathological states of the heart may drastically alter left ventricular pressure waveform shapes. Future work will include adding pathological states to the numerical heart failure model.

Controlling LVEDP at the expense of aortic pressure or inappropriate LVAD flow that could result in thrombogenesis or hemolysis also needs to be taken into consideration. Future work will involve developing an adaptive target for LVEDP that considers both arterial pressure and flow rate.

## V. Conclusion

In this study, a novel physiological control for an implantable heart pump has been developed to respond to interpatient and intrapatient variations to maintain the LVEDP in the normal range of 3 to 15 mmHg to prevent ventricle suctions and pulmonary congestions. A new algorithm was developed to detect LVEDP from pressure sensor measurement in real-time mode. Interpatient and intrapatient variations have been simulated based on sensitivity analysis, in which the most effective parameters of CVS have been determined by changing the CVS parameter between -20% to +20 of their nominal values. Tracking controller performance has been assessed via simulation of 100 different patient conditions in each of six different patient scenarios showing the preference of MFAC compared to PID control reducing the risk of suction and congestion events in patients.

## Acknowledgment

The authors would like to recognize the financial assistance provided by the National Health and Medical Research Council Centers for Research Excellence (APP1079421).